\documentclass[fleqn,usenatbib]{mnras}

\usepackage{newtxtext,newtxmath}

\usepackage[T1]{fontenc}

\DeclareRobustCommand{\VAN}[3]{#2}
\let\VANthebibliography\thebibliography
\def\thebibliography{\DeclareRobustCommand{\VAN}[3]{##3}\VANthebibliography}

\usepackage{graphicx}	
\usepackage{amsmath}	
\usepackage{color}
\usepackage[dvipsnames]{xcolor}
\usepackage{graphicx}
\usepackage{subcaption}
\usepackage{colortbl} 

\defcitealias{starforge.methods}{Paper I}
\defcitealias{starforge_jets_imf}{Paper II}

\usepackage[normalem]{ulem}
\newcommand{\msun}{\mathrm{M}_{\rm \sun}}

\newcommand{\blue}[1]{\textbf{\textcolor{blue}{#1}}}

\newcommand{\add}[1]{{\blue{#1}}}

\title[Realistic GMC initial conditions]{Less wrong: a more realistic initial condition for simulations of turbulent molecular clouds}

\author[H. B. Lane et al.]{
Henry B. Lane,$^{1,2}$\thanks{hlane17phys@gmail.com}
Michael Y. Grudi\'{c},$^{1,3}$\thanks{mgrudic@carnegiescience.edu}\thanks{NASA Hubble Fellow}
D\'avid Guszejnov,$^{4}$
Stella S. R. Offner,$^{4}$ \newauthor
Claude-Andr{\'e} Faucher-Gigu{\`e}re,$^{1}$ and
Anna L. Rosen$^{5}$
\\
$^{1}$CIERA and Department of Physics and Astronomy, Northwestern University, 1800 Sherman Ave, Evanston, IL 60201, USA\\
$^{2}$Pennsbury High School, 608 South Olds Blvd, Fairless Hills, PA 19030, USA\\
$^{3}$Carnegie Observatories, 813 Santa Barbara St, Pasadena, CA 91101, USA\\
$^{4}$Department of Astronomy, The University of Texas at Austin, TX 78712, USA \\
$^{5}$Center for Astrophysics $|$ Harvard \& Smithsonian, 60 Garden St, Cambridge, MA 02138, USA \\
}

\date{Accepted XXX. Received YYY; in original form ZZZ}

\pubyear{2021}

\begin{document}
\label{firstpage}
\pagerange{\pageref{firstpage}--\pageref{lastpage}}
\maketitle
\begin{abstract}
Simulations of isolated giant molecular clouds (GMCs) are an important tool for studying the dynamics of star formation, but their turbulent initial conditions (ICs) are uncertain. Most simulations have either initialized a velocity field with a prescribed power spectrum on a smooth density field (failing to model the full structure of turbulence) or ``stirred" turbulence with periodic boundary conditions (which may not model real GMC boundary conditions). We develop and test a new GMC simulation setup (called {\small TURBSPHERE}) that combines advantages of both approaches: we continuously stir an isolated cloud to model the energy cascade from larger scales, and use a static potential to confine the gas. The resulting cloud and surrounding envelope achieve a quasi-equilibrium state with the desired hallmarks of supersonic ISM turbulence (e.g. density PDF and a $\sim k^{-2}$ velocity power spectrum), whose bulk properties can be tuned as desired. We use the final stirred state as initial conditions for star formation simulations with self-gravity, both with and without continued driving and protostellar jet feedback, respectively. We then disentangle the respective effects of the turbulent cascade, simulation geometry, external driving, and gravity/MHD boundary conditions on the resulting star formation. Without external driving, the new setup obtains results similar to previous simple spherical cloud setups, but external driving can suppress star formation considerably in the new setup. Periodic box simulations with the same dimensions and turbulence parameters form stars significantly slower, highlighting the importance of boundary conditions and the presence or absence of a global collapse mode in the results of star formation calculations.
\end{abstract}




\section{Introduction}
\label{sec:intro}

The initial conditions (ICs) for star formation (SF) are giant molecular clouds (GMCs), which are complex objects with a highly dynamic life cycle. Their evolution is regulated by supersonic magnetohydrodynamic (MHD) turbulence, with a substructured, filamentary morphology, and have energy continuously cascading from the turbulent eddies on the galactic scale-height down to smaller scales \citep{maclow:2004,mckee_ostriker_review,hopkins:2012.excursion.set,padoan:2016.sne.driving}. GMCs form and disperse dynamically as part of the gas cycle in galaxies, exchanging mass and energy with the surrounding interstellar medium \citep[ISM;][]{hopkins:fb.ism.prop, dobbs:2013, ibanez:2016.gmc.sims,benincasa:2020.fire.gmcs,guszejnov_GMC_cosmic_evol,chevance:2020.gmc.review}. Star formation involves various nonlinear, coupled physical processes acting on a large range of physical scales, so an important tool for understanding it is numerical simulations \citep[e.g.][]{naab_ostriker_galform_review,2018NatAs...2..368F,teyssier:sf.sims.review}. However many different factors can potentially affect the outcome of SF, so the initial and boundary conditions for such simulations should be as realistic as possible to model the true complexity of GMC evolution.

An ideal solution would be to simulate the entire multi-phase gas cycle within the galaxy with GMCs forming and dispersing dynamically \citep[e.g.][]{dobbs:2008.gmcs, hopkins:fb.ism.prop,agertz:2013.new.stellar.fb.model,hu:2016}, but this is not currently feasible at the level of detail needed to capture the formation of individual protostars and the emergence of the stellar initial mass function (IMF). IMF-resolving simulations are currently practical for either {\it isolated} GMCs or dense clumps, or a presumed small ``patch" within a larger GMC, with a total gas mass limited by computational capabilities. Lacking the greater galactic context of GMC formation, such simulations must assume some simplified initial and boundary conditions, generally informed by the phenomenology of supersonic ISM turbulence.

A popular approach \citep[e.g.][]{bate2003} is to initialize a uniform-density sphere with a Gaussian random velocity field with a $\propto k^{-2}$ power spectrum \citep{dubinski:1995.turb.spectrum} to emulate turbulence (which we refer to as the {\small SPHERE} setup in this and related works --  \citealt{guszejnov_isothermal_mhd,starforge.methods,starforge_jets_imf}). Such Gaussian ICs mimic real ISM turbulence, insofar as they model its 2-point velocity statistics initially \citep[e.g.][]{larson:gmc.scalings} by construction. Due to its simplicity, the {\small SPHERE} setup remains in use today as the most widely-adopted initial condition for simulations of isolated star-forming gas clouds and cores, with various variations in the initial density profile employed \citep{Girichidis_2011_isoT_sim,rosen_2016_massive_sf,Lee_Hennebelle_2018_IC, Rosen2019a}, and typically with a uniform initial magnetic field when magnetic fields are included \citep[e.g.][]{price_bate_2007_mhd_sf,offner:2017.jets,Rosen2020a}. Simulations of this type have been shown to reproduce certain aspects of GMCs found in self-consistent global galaxy simulations, in particular the density PDF once evolved from the initial uniform-density state \citep{2015MNRAS.446L..46R}.

However, fully-developed turbulence is nonlinear with high-order correlations between the density, velocity, and magnetic field. Such correlations are not captured by enforcing a certain initial velocity power spectrum.\footnote{Another notable, prior approach to Gaussian ICs is found in \citet{klessen:2000.cloud.collapse}, who used Lagrangian perturbation theory to generate an initial density structure consistent with a Gaussian random velocity field with a prescribed $\propto k^{-2}$ power spectrum, with periodic boundary conditions. This is nominally more realistic than assuming uniform density, but perturbation theory still cannot capture fully-developed turbulence.} Even if the power spectrum is tuned to reproduce observations, this does not imply a true turbulent {\it cascade} of energy from large scales down to the dissipation scale. Without ongoing energy injection, smaller eddies decay on their shorter turnover timescale $\tau_{\rm edd} \sim \lambda /v\left(\lambda\right) \sim \lambda^{0.5}$ \citep{gammie.ostriker:1996.mhd.dissipation, maclow:1999.turbulence}, changing the shape of the power spectrum over time. Lastly, the lack of material surrounding the cloud is not clearly realistic, as there would likely be some continuity between the material in the ``cloud" and the larger-scale galactic gas flow that formed it. Therefore a more realistic IC is desirable.

Other simulations have modeled the initial turbulent state by "stirring" turbulence in a periodic box for a certain length of time \citep[e.g.][]{maclow:1999.turbulence}, then switching on gravity to allow gravitational collapse and the formation of cores and/or stars from a self-consistent initial turbulent state, optionally with continued turbulent driving \citep{li:2004.mhd.turb.sf}. With continued driving, a true turbulent cascade with a steady power spectrum emerges (although it is still artificially truncated by numerical dissipation process at the resolution scale, \citealt{schmidt:2004.bottleneck}). Although it is still artificial to "switch on" gravity instantaneously, such simulations (denoted {\small BOX} simulations in this work) are in a sense more realistic in both their initial state (determined by the emergent dynamics of turbulence), and in how they can model the ongoing injection of turbulent energy from larger scales not captured in the dynamic range of the simulation. 

{\small BOX}-type simulations have sometimes arrived at different conclusions than equivalent {\small SPHERE} simulations for many important aspects of SF, including the rate of star formation, the shape of the IMF, the physical properties of cores and disks, stellar multiplicity, and the clustering of star formation \citep{Offner_2008,krumholz:2012.orion.rhd,federrath_klessen_2012,myers:2014.feedback,federrath_2015_inefficient_sf,guszejnov_isothermal_mhd,starforge_jets_imf}. Although such differences might be interpreted physically as the effect of more-realistic, self-consistent turbulence and/or driving, it is important to note that {\small BOX} models have differed from {\small SPHERE} models in multiple ways:
\begin{enumerate}
    \item Self-consistency of the initial turbulent state: {\small SPHERE} runs are generally run from approximate Gaussian initial conditions, while {\small BOX} runs begin with a self-consistent turbulent state.
    \item Driving versus decay: {\small SPHERE} runs are typically run in a decaying configuration, whereas {\small BOX} runs often continue driving after gravity is switched on to model the cascade.
    \item (Magneto-) hydrodynamic boundary conditions: {\small SPHERE} runs generally adopt vacuum boundary conditions or periodic boundary conditions with a box size much larger than the cloud \citep[e.g.][]{price_bate_2008_mhd_cluster}, whereas {\small BOX} runs have periodic boundary conditions in a configuration where the cloud mass fills the whole box uniformly on average.
    \item Gravitational boundary conditions: the solution to the Poisson equation in {\small SPHERE} simulations results in a confining central gravitational acceleration that is absent in a periodic {\small BOX} simulation. For a given set of turbulent conditions, {\small BOX} simulations can thus have an order-of-magnitude less gravitational energy  \citep{federrath_klessen_2012}.
\end{enumerate}

The boundary conditions of {\small BOX} setups may be problematic once feedback is introduced into the simulation: material that should physically be expelled from the cloud will instead run laps around the box \citep{2007ApJ...662..395N}, and even if feedback excites the box to a highly-turbulent state the system cannot expand as it would according to the virial theorem. Mass loss and disruption by feedback are a key phase of GMC evolution likely determining the end of star formation, but a GMC-sized periodic box setup cannot model such processes fully.

Moreover, a periodic solution to the Poisson equation freezes out the global {\it gravitational} mode, i.e. there is zero gravitational field in the absence of density perturbations, where physically the boundary conditions for any centrally concentrated gas structure would have a global confining potential. This global potential could potentially affect gas and stellar dynamics, the IMF (e.g. if there is a tendency for massive stars to sink to the global potential minimum and accrete gas, \citealt{bonnell:2001.competitive.accretion}), and star cluster structure and assembly (neighbouring subclusters are more likely to merge in a global confining potential, e.g. \citealt{bonnell:2003.hierarchical,grudic:2017}). Practically by definition, GMCs are overdensities with respect to the surrounding ISM, so it could be more realistic to account for the confining potential that generically arises.

Thus we desire a simulation setup that has the best of both worlds: a setup that is localized in space with outflow or low-density boundary conditions like the {\small SPHERE} setup, but has the fully-developed initial turbulence -- and optional continued energy injection -- of the {\small BOX} setup. In this work, we present a new type of simulation setup for GMCs, called {\small TURBSPHERE}, that has these properties. In \S\ref{sec:methods} we describe our simulation code, our implementations of the {\small SPHERE} and {\small BOX} setups, and the new {\small TURBSPHERE} setup, and outline two suites of numerical simulations: one to investigate the behavior of the {\small TURBSPHERE} turbulence stirring procedure, and the other including gas self-gravity to investigate the resulting star formation. In \S\ref{sec:results} we present the results of our simulation suites for different cloud parameters. In \S\ref{sec:discussion} we discuss the implications of our results regarding the roles of turbulence, continued driving, initial conditions, and boundary conditions in star formation simulations, and in \S\ref{sec:conclusion} we summarize our main findings.

\section{Methods}
\begin{figure*}
 \centering
    \includegraphics[width=\textwidth]{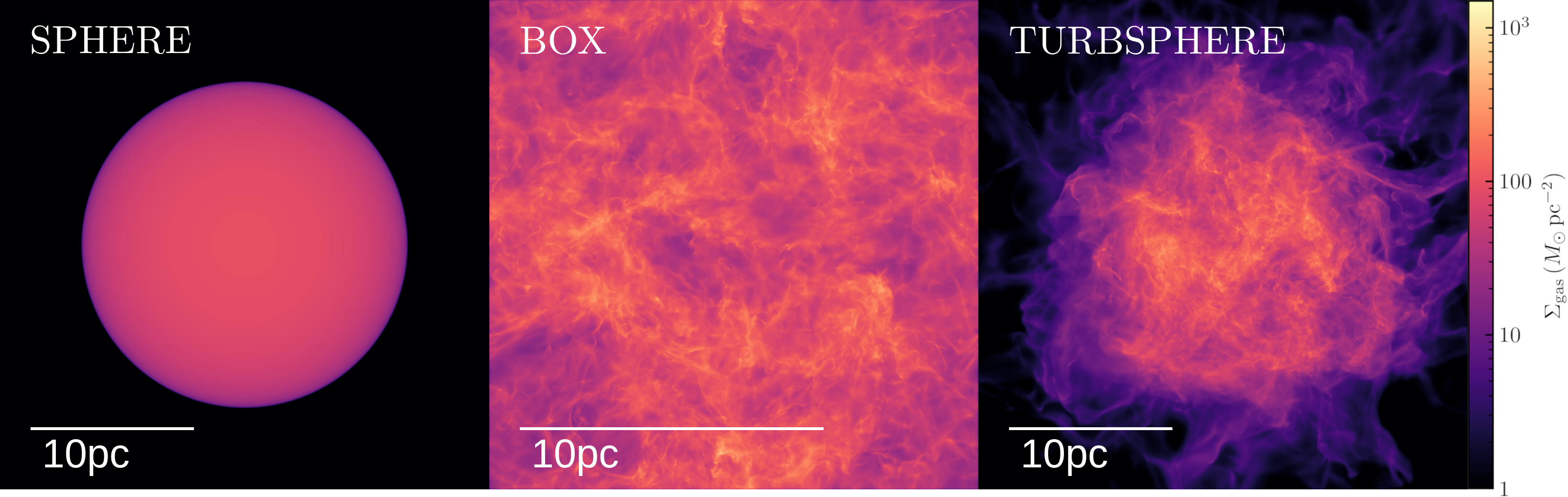}

    \caption{\label{fig:multi_panel1} Surface density maps of our initial condition setups. {\small SPHERE} (left) is a uniform sphere with a pre-generated, pseudo-turbulent $\propto k^{-2}$ velocity power spectrum. {\small BOX} (middle) shows the periodic-box setup, where fully-realized, self-consistent turbulence is simulated before self-gravity is enabled. {\small TURBSPHERE} (right) features a roughly spherical cloud with self-consistent turbulence resulting from stirring the gas while confining it with an ad-hoc potential.}
\end{figure*}

\label{sec:methods}

\subsection{Simulation code}
We utilize the STar FORmation in Gaseous Environments ({\small STARFORGE}) numerical framework implemented within the {\small GIZMO} code to perform our simulations. A full discussion and presentation of the {\small STARFORGE} framework, along with tests and algorithm details, is available in \citet{starforge.methods} (hereafter \citetalias{starforge.methods}). Therefore, we only summarize the physics modules utilized below.

\subsubsection{Core physics}
We use the quasi-Lagrangian meshless finite-mass (MFM) method for magnetohydrodynamics (MHD) from \citet{hopkins:gizmo.mhd}. We assume ideal MHD with a constrained-gradient scheme that ensures $\nabla \cdot B = 0$ to high precision \citep{hopkins:2016.divb}. For simplicity, all simulations in this work assume an isothermal equation of state $P = \rho c_s^2$, where $c_{\rm s} = 0.2\, \rm km\,s^{-1}$ is the typical isothermal sound speed in $\sim$10K molecular gas.

Gravity is solved with an improved version of the parallel Barnes–Hut tree solver from \citet{springel:2005} combined with high-order integration of sink particle trajectories to accurately track multiple-star systems \citepalias{starforge.methods}. We optimize the gravity calculation for gas using the \citet{grudic:adaptive} adaptive force updating scheme. Softening is fully adaptive for gas cells, while sink particles (representing individual (proto)stars) adopt a softening radius of $21 \rm AU$. Sinks follow the {\small STARFORGE} formation and accretion algorithm derived from that of \citet{bate.sph.fragment}, requiring multiple checks to form a sink particle, which accretes other Lagrangian gas cells based on criteria accounting for magnetic, kinetic and gravitational energies and angular momentum.  

Upon forming, sinks follow the protostellar evolution model described in \citet{offner:star.evo.2009}, treating the stars as collapsing polytropes with distinct phases. These phases ("pre-collapse", "no burning", "core deuterium burning at fixed temperature", "core deuterium burning at variable temperature", "shell deuterium burning" and "main sequence") allow for dynamic evolution of the stellar properties, in particular the protostellar radius that sets the jet velocity.

\subsubsection{Protostellar jets}
Protostars eject a significant portion of their accreting mass in bipolar jets  \citep{bally:2016.jets.review,pudritz:2019.jets.review}. These jets have been shown to suppress star formation globally and locally, significantly reducing the overall SF efficiency and the mass-scale of individual stars
\citep{hansen:2012.jets,federrath_2014_jets, Rosen2020a, starforge_jets_imf,mathew:2021.imf.jets}. As these effects can be so important, we account for protostellar jet feedback in a subset of our star formation simulations. We use the parametrization of \citet{Cunningham.2011.momentum}, wherein a fraction $f_{\rm w}$ of the material accreting near the protostar is launched in a bipolar collimated jets along the sink angular momentum axis, with velocity
\begin{equation}
    v_{\rm jet} = f_{\rm K} \sqrt{\frac{GM_{\star}}{R_{\star}}},
\end{equation}
where $f_{\rm K}$ represents the fraction of Keplerian velocity at the protostellar radius $R_{\star}$. We assume $f_{\rm w}$ = $f_{\rm K}$ = 0.3, setting the momentum loading parameter $f_{\rm w}f_{\rm K}$ in the range inferred from observations and in line with other works \citep{Cunningham.2011.momentum,offner:2017.jets}. Jet feedback is implemented numerically by ``spawning" new Lagrangian gas cells near the sink particle. Further in depth discussion on the protostellar jets implementation can be found in the {\small STARFORGE} methods paper \citepalias{starforge.methods}.

\subsubsection{Turbulence driving}
For simulations with external driving, we use the turbulence driving scheme introduced in \citet{schmidt:2004.bottleneck}. We stir gas throughout the box with a body force considering of Fourier modes corresponding to wavelengths ranging from 0.25 - 1 times the size scale of the simulation (cloud radius for isolated clouds, box size for periodic {\small BOX }ICs), with a peak amplitude at $\lambda = 0.5\, \rm R_{\rm 0}$ by default. These modes are generated in Fourier-space as an Ornstein-Uhlenbeck process with a coherence timescale of roughly 1 global free-fall time ($\tau = 3 \rm Myr$). The compressive part of the acceleration is projected out via a Helmholtz decomposition in Fourier space so that the driving is purely solenoidal \citep{federrath:2008.driving}, modeling the effect of large-scale supernova driving on GMC scales \citep{padoan:2016.sne.driving}.

\label{sec:driving_methods}
\subsection{Initial conditions and setups}

In this work we perform simulations with three different choices of initial and boundary conditions, comparing {\small SPHERE} and {\small BOX} setups (which are commonly employed in star formation studies) to our new {\small TURBSPHERE} setup, which we detail below. Figure~\ref{fig:multi_panel1} shows density projections of each simulation set-up at initialisation, which we describe in following sections. In all simulations we adopt a mass resolution $\Delta m=10^{-3}\msun$, resolving a $M_{\rm 0} = 2\times 10^4 M_\odot$ cloud with $2\times 10^7$ equal-mass gas cells. 

\subsubsection{{\small SPHERE}}
Our {\small SPHERE} IC
 consists of a spherical cloud of uniform density with mass $M_{\rm 0} = 2\times 10^4 \msun$ and radius $R_{\rm 0} = 10$ pc, positioned in the center of a periodic box of side length 10$R_{\rm 0}$. The cloud is enveloped by a diffuse medium with a 1/1000 density contrast.
The initial magnetic field is uniform, with its strength  $B_{\rm 0}$ set by the chosen dimensionless mass-to-flux ratio $\mu_{\rm 0}$, the ratio of magnetic flux through the cloud to the critical flux determined in \citep{Mouschovias_Spitzer_1976_magnetic_collapse}:
\begin{equation}
    \mu_{0} =  0.4 \frac{G^{1/2} M_{\rm 0}}{\uppi R_{\rm 0}^2 B_{\rm 0}} = 0.42 \sqrt{\frac{|E_{\rm grav}|}{E_{\rm mag}}},
\end{equation}
where $E_{\rm grav}$ and $E_{\rm mag}$ are the gravitational and magnetic energy of the gas within the cloud, and $\mu_{\rm 0}=4.2$ is our fiducial value consistent with the observed magnetic field strengths of dense molecular gas \citep{Crutcher2010a}. Finally, to mimic the velocity correlations of turbulence, an initial velocity field is initialized, where a Gaussian random field (with a power spectrum of $P_{\rm v}\left(k\right) \propto k^{-2}$) is generated on a Cartesian grid and interpolated onto the cell positions. The velocity is normalized to achieve an initial virial parameter $\alpha_{\rm turb}=5 R_{\rm 0} \mathcal{M}^2 c_{\rm s}^2/\left(3 G M_{\rm 0}\right)=2$. However, $\alpha_{\rm turb} \sim 1$ during most of the star formation due to the decay of turbulence.

\subsubsection{{\small BOX}}
In the {\small BOX} setup, gas cells are initially deposited on a uniform-density glass configuration in a periodic box with zero velocity. Similar to other runs in this suite, {\small BOX} runs have a mass $M_{\rm 0} = 2\times 10^4 \msun$ and have a size set to have the same volume-density ($L_{\rm box}$ = 16.1pc) as {\small SPHERE} runs. During initialization, gravity is disabled, and the periodic box is continually stirred by an injection of energy with a predetermined power spectrum as described in \S\ref{sec:driving_methods}, tuned to achieve an RMS Mach number $\mathcal{M} = \sqrt{\langle v^2 /c_{\rm s}^2 \rangle} \sim 10$. This is similar to the Mach number achieved in our fiducial {\small TURBSPHERE} setup and the Mach number in our {\small SPHERE} setup at the time that most stars form. The {\small BOX} IC initialization transforms an initial uniform magnetic field, via self-consistent evolution, to a magnetic field that mimics MHD turbulence. 

After five free-fall times, gravity is enabled. Thus, simulations using the {\small BOX} IC start with self-consistent MHD turbulence, wherein the resulting density distribution is nearly lognormal \citep{vazquez-semadeni:1994.turb.density.pdf,padoan:1997.density.pdf} and the density, velocity, and magnetic fields have non-linear correlations.

\subsubsection{{\small TURBSPHERE}}
{\small TURBSPHERE} initial conditions are generated by invoking a {\small SPHERE} initial condition setup, introducing an analytic potential, and enabling driving to initialize a self-consistent turbulent state. The initial magnetic field is uniform with $\mu_{0}=4.2$ by default.

The turbulence initialization procedure uses an analytic potential -- as opposed to the full self-gravity solution -- because we wish to initialize the turbulent state self-consistently (as is also done in our BOX setup) without allowing stars to form at first \add{}. We have also experimented with two alternate approaches for modelling gravity while preventing SF: using a large gravitational softening length ($\epsilon \gtrsim 1 \rm pc$), and making the gravity solution spherically symmetric while evolving the MHD equations in 3D. Using a large softening still allowed the spontaneous formation of bound structures and runaway collapse, while the spherically-symmetric solver led to large, unstable swings in the cloud properties and either runaway collapse or cloud dispersal. We seek an initial condition close to some notion of equilibrium, so we adopt a static potential.

Inside the nominal cloud radius, $R_{\rm 0}$, the analytic gravitational field is that of a homogeneous sphere matching the nominal cloud properties ($R_{\rm 0} = 10$ pc, and mass $M_{\rm well} = 2\times 10^4  \msun$, unless we explicitly vary $M_{\rm well}$ to test the sensitivity of this parameter). Within $r<R_{\rm 0}$, the radial gravitational field is
\begin{equation}
    g_{\rm r} = - \frac{G M_{\rm well}r}{{R_{\rm 0}}^3}.
\end{equation}
Outside $R_{\rm 0}$, we add an additional component modeling the field of mass halo with a $\rho \propto r^{-3}$ density profile, matched in density to the uniform sphere at $R=R_{\rm 0}$. Hence for $r > R_{\rm 0}$, the field is given by
\begin{equation}
    g_{\rm r} = - \frac{GM_{\rm well}\left(1 + 3 \ln \left(\frac{r}{R_{0}}\right)\right)}{r^2}.
\end{equation}
The inclusion of the additional halo component gives the potential infinite escape speed (to prevent mass loss) without imposing a sharp confining barrier, because $\rho \propto r^{-3}$ is the steepest power-law mass model with infinite escape speed. During the turbulence initialization, gas cells are only subject to the analytic potential well and do not self-interact via gravity.

A driving force is applied as described in \S\ref{sec:driving_methods} as the simulation progresses, similar to the BOX case. The driving scheme is identical to that of {\small BOX} runs, and we tune the driving strength to achieve a nominal $\mathcal{M} \sim 10$, which is chosen to match {\small SPHERE} runs at the time that SF begins in those runs. 
We run the initialization until the gas reaches a quasi-equilibrium state, as we will demonstrate in \S\ref{sec:results}. 

\subsection{Simulation suites}
We run two suites of simulations in order to map the parameter space of possible {\small TURBSPHERE} initial conditions and determine the effects on subsequent star formation.

The first suite aims to investigate how the {\small TURBSPHERE} initialization setup behaves qualitatively, and how varying its free parameters affects the resulting dynamical equilibrium. We perform 8 {\small TURBSPHERE} initialization runs, varying the scaling factor of the forcing strength $\tilde{f}$, the scaling factor of the wavelength range of our driving scheme $\tilde{\lambda}$, the strength of the analytic gravity well, and the mass-to-flux ratio $\mu_{\rm 0}$. All runs are listed in Table \ref{table:stirring_runs} and discussed in full in \S\ref{sec:stirring_runs_sec}.
\begin{table*}
    \centering
    \renewcommand{\arraystretch}{1.5}
    \setlength\tabcolsep{3.5pt}
    \begin{tabular}{c c c c|c c c c c}
    \hline 
    Forcing strength & Forcing wavelength range ($R_{\rm 0}$)  & $M_{\rm well}$ ($M_{\rm 0}$) & $\mu_{\rm 0}$ & $\mathcal{M}_{\rm RMS}$ & $R_{\rm 50}/R_{\rm 0}$ & $R_{\rm RMS}/R_{\rm 0}$ & $E_{\rm mag}/E_{\rm kin}$ & $\alpha_{\rm turb} = 2 E_{\rm kin}/|E_{\rm self-gravity}|$ \\
    \hline
    1 & $\left[1,0.25\right]$  & 1 & 4.2 & $10.9 \pm 0.3$ & $0.77 \pm 0.05$ & $0.88 \pm 0.04$ & $0.31 \pm 0.03$ & $1.0 \pm 0.1$\\
    \hline
    0.7 & $\left[1,0.25\right]$  & 1 & 4.2 & $9.5 \pm 0.5$ & $0.63 \pm 0.06$ & $0.72 \pm 0.05$ & $0.27 \pm 0.03$ & $0.6 \pm 0.1$\\
    \hline
    1.4 & $\left[1,0.25\right]$  & 1 & 4.2 & $12.1 \pm 0.4$ & $0.90 \pm 0.04$ & $1.09 \pm 0.06$ & $0.36 \pm 0.03$ & $1.7 \pm 0.1$\\
    \hline
    1 & $\left[1,0.25\right]$  & 0.5 & 4.2 & $9.8 \pm 0.2$ & $1.06 \pm 0.05$ & $1.34 \pm 0.03$ & $0.38 \pm 0.03$ & $1.7 \pm 0.1$\\
    \hline
    1 & $\left[1,0.25\right]$  & 2 & 4.2 & $11.5 \pm 0.7$ & $0.51 \pm 0.04$ & $0.58 \pm 0.02$ & $0.25 \pm 0.04$ & $0.62 \pm 0.09$\\
    \hline
    1 & $\left[2,0.50\right]$  & 1 & 4.2 & $14.3 \pm 0.4$ & $1.10 \pm 0.09$ & $1.64 \pm 0.04$ & $0.22 \pm 0.02$ & $3.6 \pm 0.3$\\
    \hline
    1 & $\left[1,0.25\right]$  & 1 & 1.3 & $10.5 \pm 0.4$ & $0.73 \pm 0.04$ & $0.84 \pm 0.02$ & $0.66 \pm 0.06$ & $0.89 \pm 0.08$\\
    \hline
    1 & $\left[1,0.25\right]$  & 1 & 0.42 & $10.4 \pm 0. 2$ & $0.89 \pm 0.02$ & $1.05 \pm 0.01$ & $3.2 \pm 0.1$ & $1.07 \pm 0.03$\\
    \hline
    \end{tabular}
    \caption{List of {\small TURBSPHERE} turbulence initialization run parameters and results. Columns give (1) the normalization of the turbulent forcing relative to the fiducial run, (2) the range of wavelengths of driving modes in units of the initial cloud radius $R_{\rm 0}$, (3) the mass sourcing the analytic gravity well in units of the nominal cloud mass $M_{\rm 0}$, (4) the initial cloud mass-to-flux ratio, and the final equilibrium values of the cloud 3D RMS Mach number $\mathcal{M}_{\rm RMS}$, half-mass radius $R_{\rm 50}$, RMS distance from center of mass $R_{\rm RMS}$, ratio of magnetic to kinetic energy in the cloud, and cloud virial parameter accounting for the cloud self-gravity $\alpha_{\rm turb}$ (Columns 5-9 respectively). Equilibrium values are measured over the last 10 crossing times, and we also quote their $\pm \sigma$ variance over this interval.}
    \label{table:stirring_runs}
\end{table*}

We then run eight different star formation simulations to determine how simulations from the new {\small TURBSPHERE} setup behave, and to identify the effects of protostellar jets, the turbulent cascade, simulation geometry, external driving, and gravity/MHD boundary conditions on the subsequent star formation statistics. We perform two {\small SPHERE} runs with and without protostellar jets, two driven {\small BOX} runs with and without jets, and four {\small TURBSPHERE} runs, switching driving and protostellar jets in each possible combination. We list the initial conditions and parameters of the star formation runs in Table \ref{table:runs_dir} and discuss their results in \S\ref{sec:star_formation_runs}. 

\begin{table*}
    \centering
    \renewcommand{\arraystretch}{1.2}
    \begin{tabular}{l|c|c|c|c|c|c|c|c|c}
    \hline 
    Star formation simulations & Method & Driving & Jets & $M_{0} (\msun)$  & $\mu_0$ & $\mathcal{M}_{\rm RMS}$ & Size (pc) & $\alpha_{\rm turb}$ & $\rm SFE_{\rm final}$\\
    \hline
    {\small TURBSPHERE} & {\small TURBSPHERE} & No & No & $2\times 10^4$ & 1.1 & 11.2 & 7.7 & 1.1 & 0.22\\
    \hline
    {\small TURBSPHERE+Jets} & {\small TURBSPHERE} & No & Yes & $2\times 10^4$ & 1.1 & 11.2 & 7.7 & 1.1 & 0.11\\
    \hline
    {\small TURBSPHERE+Driving} & {\small TURBSPHERE} & Yes & No & $2\times 10^4$ & 1.1 & 11.2 & 7.7 & 1.1 & 0.15\\
    \hline
    {\small TURBSPHERE+Driving+Jets} & {\small TURBSPHERE} & Yes & Yes & $2\times 10^4$ & 1.1 & 11.2 & 7.7 & 1.1 & 0.06\\
    \hline
    {\small BOX+Driving} & {\small BOX} & Yes & No & $2\times 10^4$ & $4.2$ & $11.4$ & 10 & 22 & 0.07\\
    \hline
    {\small BOX+Driving+Jets} & {\small BOX} & Yes & Yes & $2\times 10^4$ & $4.2$ & $11.4$ & 10 & 22 & 0.01\\
    \hline
    {\small SPHERE} & {\small SPHERE} & No & No & $2\times 10^4$ & 4.2 & 16.1 & 7.9 & 2.0 & 0.29\\
    \hline
    {\small SPHERE+Jets} & {\small SPHERE} & No & Yes & $2\times 10^4$ & 4.2 & 16.1 & 7.9 & 2.0 & 0.11\\
    \hline
    \end{tabular}
    \caption{List of star formation runs with their initial conditions and parameters. Columns give (1) the star formation simulation name, (2) the initial condition method used, (3) if driving is enabled, (4) if protostellar jets are enabled, (5) the initial stellar mass $(M_0)$, (6) the initial mass-to-flux ratio ($\mu_0$), (7) the 3D RMS Mach number ($\mathcal{M}_{\rm RMS}$), (8) the size of the initial condition ($R_{50}$ for {\small TURBSPHERE} \& {\small SPHERE} runs, $L_{\rm box}$ for {\small BOX runs}), (9) the initial virial parameter, (10) and the final star formation efficiency, $\rm SFE_{\rm final}$.} 
    \label{table:runs_dir}
\end{table*}

\section{Results}
\label{sec:results}

\begin{figure*}
 \centering
    \includegraphics[width=\textwidth]{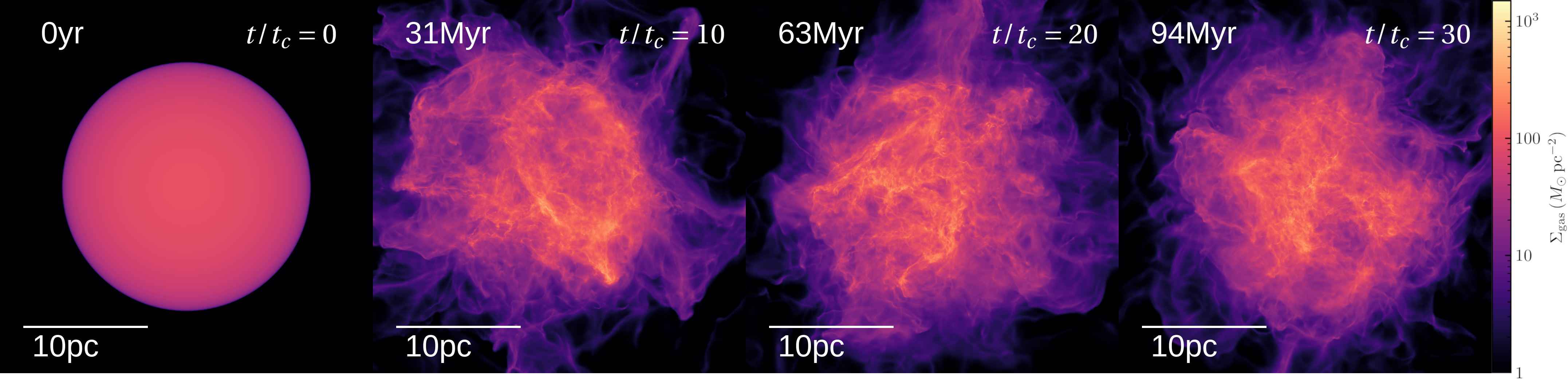}

    \caption{\label{fig:multi_panel} The evolution of our fiducial {\small TURBSPHERE} run, taken at intervals every 10 crossing times. After quasi-equilibrium is established, the cloud continues to evolve and sustain realistic turbulence through large-scale energy injection via driving.} 
\end{figure*}

\begin{figure}
    \centering
    \includegraphics[width=\columnwidth]{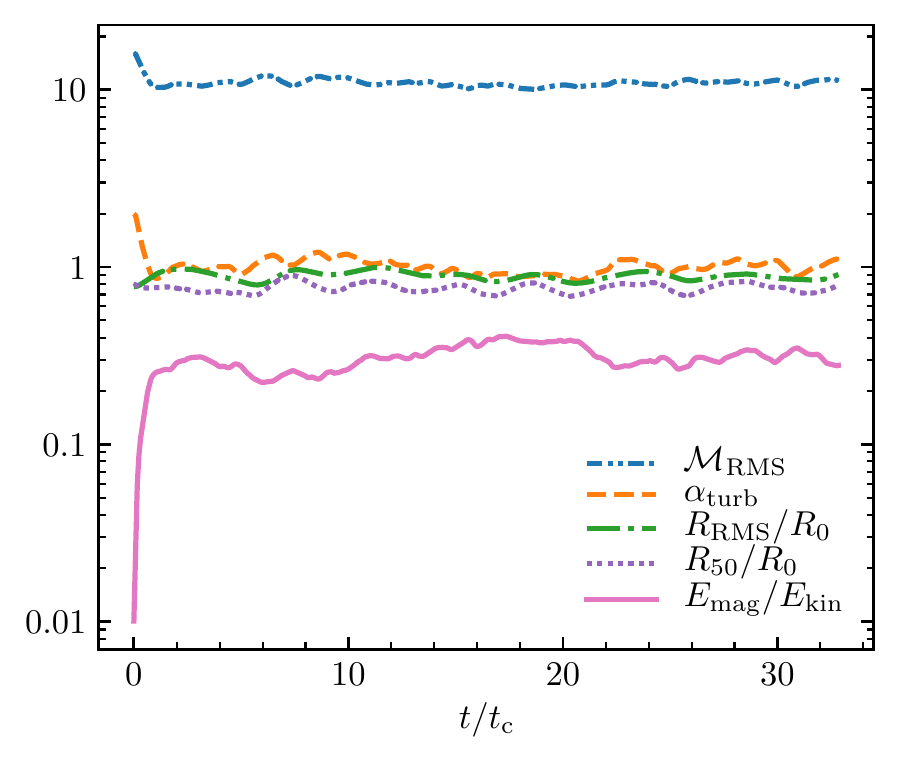}\vspace{-6mm}
    \caption{Evolution of our fiducial {\small TURBSPHERE} initialization run, plotting the 3D RMS Mach number $\mathcal{M}_{\rm RMS}$, half-mass radius $R_{\rm 50}$, RMS distance from center of mass $R_{\rm RMS}$, ratio of magnetic to turbulent energy $E_{\rm mag}/E_{\rm turb}$ and virial parameter $\alpha_{\rm turb}$ as a function of time in units of the crossing time $t_{\rm cross} = R_{\rm 0}/\left(c_{\rm s} \mathcal{M}_{\rm RMS}\right)$. All of these quantities approach equilibrium after 2-3 crossing times.} 
    \label{fig:evolution}
\end{figure}
\begin{figure}
    \centering
    \includegraphics[width=\columnwidth]{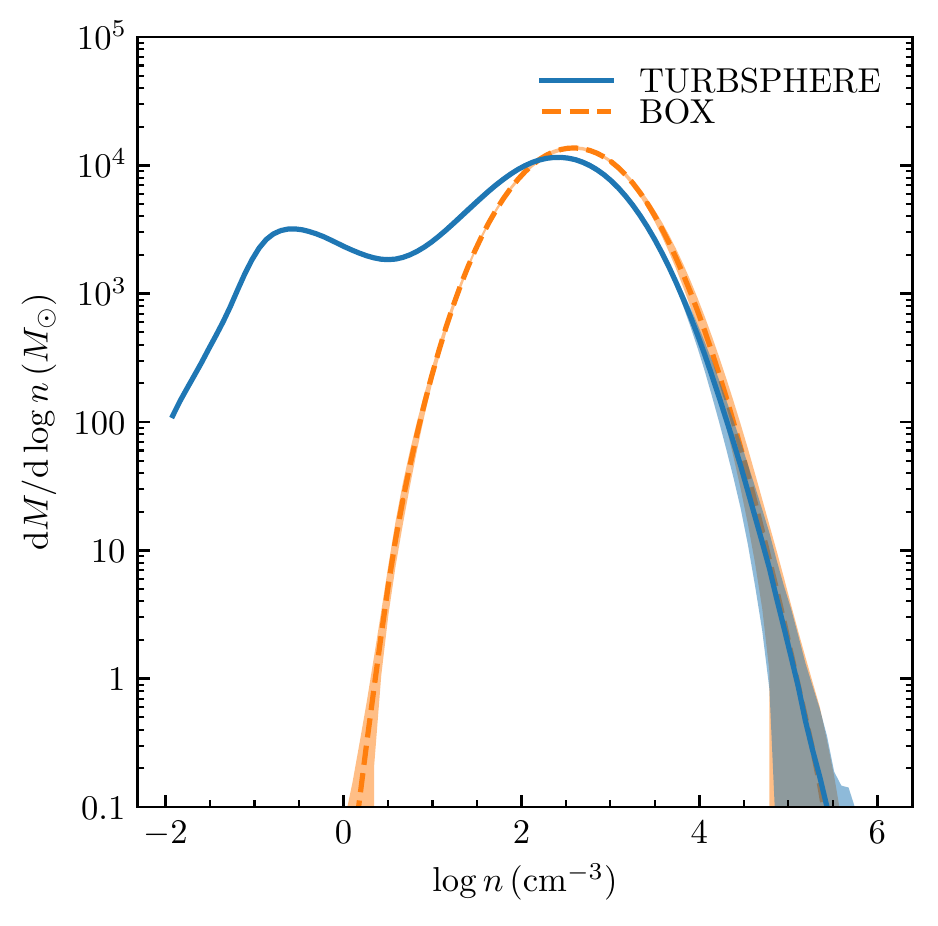}\vspace{-5mm}
    \caption{Mass-weighted density PDF in the fiducial stirring run, averaged over the final 3 crossing times with shaded interval indicating the standard deviation during this time due to intermittency. For comparison we also plot the density PDF from a $\mathcal{M} \sim 10$ driven-turbulence run in a periodic box with roughly equal volume-averaged density, averaged over the final 3 crossing times.} 
    \label{fig:rhopdf}
\end{figure}

\subsection{{\small TURBSPHERE} stirring runs}
From the initial {\small SPHERE} setup state, the {\small TURBSPHERE} initialization runs continually inject energy into the system via the driving scheme discussed in \S\ref{sec:driving_methods} through their entire duration, until quasi-equilibrium is established. To ensure that this occurs, we ran {\small TURBSPHERE} initialization runs for $\sim 30 t_{\rm c}$, where
\begin{equation}
    t_{\rm c} = \frac{R_0}{c_{\rm s} \mathcal{M}}
\end{equation}
is the turbulent crossing time. While we run the simulations for several tens of $t_{\rm c}$, an apparent quasi-equilibrium is established in a similar number of crossing times to {\small BOX} simulations ($2 \lessapprox t/t_{\rm c} \lessapprox 3$, as found in previous works, e.g. \citealt{li:2004.mhd.turb.sf}).

The evolution of the cloud over many crossing times can be seen in Figure \ref{fig:multi_panel}, where the fiducial {\small TURBSPHERE} stirring run retains a relatively spherical shape, while a halo of diffuse material (made primarily of the original diffuse envelope from the {\small SPHERE} initial conditions and ejected matter) persists around the cloud. The cloud appears to be in quasi-equilibrium, with its macroscopic properties remaining steady over time.

Analysis of the system's macroscopic properties, shown in Figure \ref{fig:evolution}, reveals that the size of the cloud is relatively stable throughout the entire evolution after $\sim 2$ crossing times, as evident by the stability of the ratio of the mass-weighted RMS and median radii $R_{\rm RMS}$ and $R_{\rm 50}$ to the initial cloud radius $R_{\rm 0}$. We also plot the virial parameter $\alpha_{\rm turb}$ (neglecting the imposed potential well and accounting for the binding energy the gas would have if self-gravity were enabled), the 3D RMS Mach number, and the ratio of magnetic to kinetic energy. 

Despite the continual injection of energy into the cloud, all quantities of interest remain relatively stable after equilibrium is established due to the ongoing dissipation of turbulence in shocks \citep{maclow:1999.turbulence} and the confinement of the central cloud by the ad-hoc potential.
\label{sec:stirring_runs_sec}

\subsubsection{Turbulence statistics}
To investigate the properties of the turbulence that develops in {\small TURBSPHERE} runs, we use our {\small BOX} initialization run as a reference, as the properties of MHD turbulence in this type of setup have been documented extensively \citep{maclow:1999.turbulence,li:2004.mhd.turb.sf,federrath:2008.driving}. The stirred {\small TURBSPHERE} cloud contains clumps and filaments with a morphology similar to those found in {\small BOX} clouds, as visually apparent in Figure \ref{fig:multi_panel1}. We compare the mass-weighted density probability distribution functions (PDFs) 
of the fiducial stirring runs of both {\small BOX} and {\small TURBSPHERE} setups in Figure \ref{fig:rhopdf}. 

Most of the mass in both PDFs is found in an approximately log-normal component, as expected generically in isothermal supersonic turbulence \citep{vazquez-semadeni:1994.turb.density.pdf,padoan:1999.density.pdf}. The additional low-density peak found in the {\small TURBSPHERE} density PDF is the diffuse, box-filling medium surrounding the cloud, and between the cloud and diffuse medium densities there is material in a halo surrounding the cloud. The {\small TURBSPHERE} and {\small BOX} density PDFs agree best at the highest densities, to the point of being practically indistinguishable in Figure \ref{fig:rhopdf}. This suggests that the two setups have overall similar density structure in the central regions of the cloud.

\begin{figure}
    \centering
    \includegraphics[width=\columnwidth]{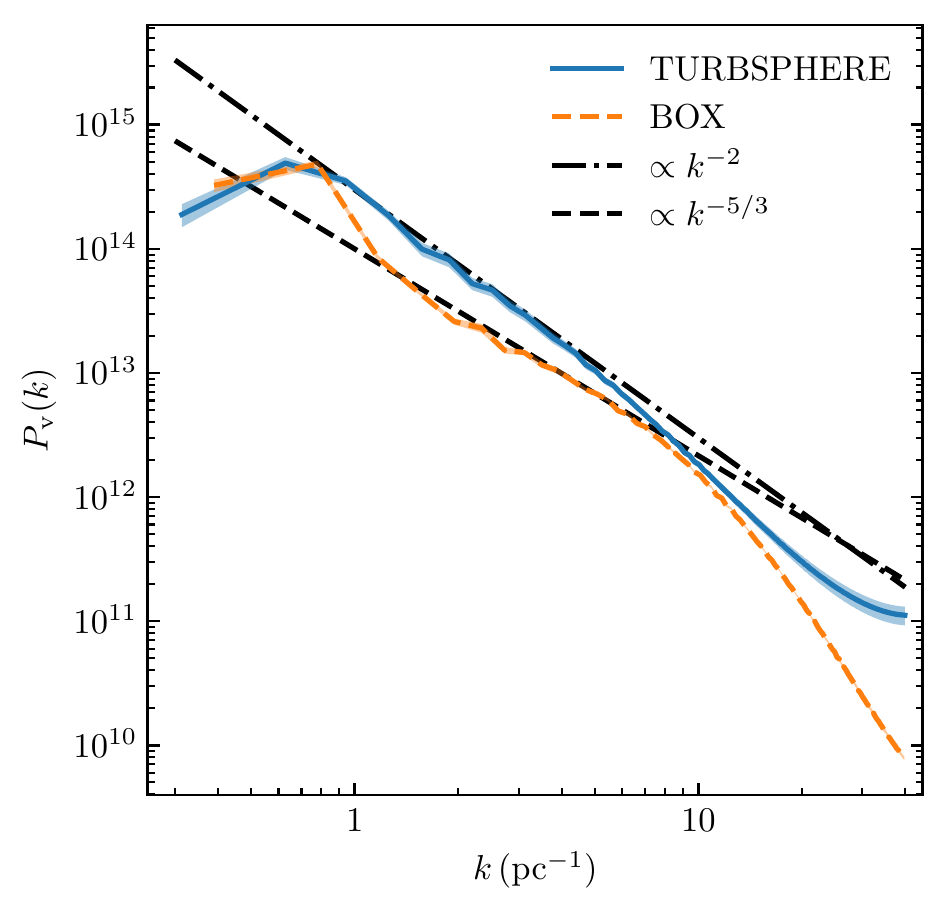}
    \vspace{-8mm}
    \caption{The velocity power spectrum of the the fiducial TURBSPHERE run, expressed as a power spectrum of the velocity field for given physical wavenumbers $k (\rm pc^{-1})$, compared to that of the equivalent BOX run and a $\propto k^{-2}$ power law. We plot the average statistics of the last five crossing times and their standard deviations (shaded regions).}
    \label{fig:powerspec_v}
\end{figure}

In Figure \ref{fig:powerspec_v}, we show the velocity power spectrum $P_{\rm v}(k)$ in the reference {\small TURBSPHERE} and {\small BOX} runs when turbulence is fully developed. Note that we expect a ``bottleneck" effect in the power spectrum on scales comparable to the mean cell spacing $\Delta x \sim 0.1 \rm pc$ where the numerical method imposes an artificial steepining of the power spectrum \citep{schmidt:2004.bottleneck,padoan:2007.mhd.cores}, and based on previous tests of supersonic turbulence in {\small GIZMO} at comparable resolution \citep{hopkins:gizmo} we expect this to occur at a physical angular wavenumber of $k \sim 100/L_{\rm box}$  in our simulations. Such steepening is evident in both the {\small TURBSPHERE} and {\small BOX} power spectra for $k \gtrsim 10 \rm pc^{-1}$, so the physical results are at wavenumbers lower than this. In this range, the {\small BOX} run exhibits a power-law slope somewhat shallower than $\propto k^{-2}$ expected for unmagnetized supersonic turbulence, and closer to the $\propto k^{-5/3}$ law found in other similar supersonic MHD turbulence simulations \citep{li:2004.mhd.turb.sf}. The {\small TURBSPHERE} setup has a somewhat steeper spectral index of $\sim -1.8$. This indicates that the choice of boundary conditions and confining potential can generally affect the velocity power spectrum measured in MHD turbulence simulations, even when controlling for $\mathcal{M}$, $\mu_{\rm 0}$, and the driving scale. Both setups' power spectral indices lie within the range of values that have been inferred in real GMCs \citep{brunt:2002.gmc.powerspectrum, padoan:2003.gmc.structfunc,padoan:2006.perseus.power.spectrum}.

\begin{figure}
    \centering
    \includegraphics[width=\columnwidth]{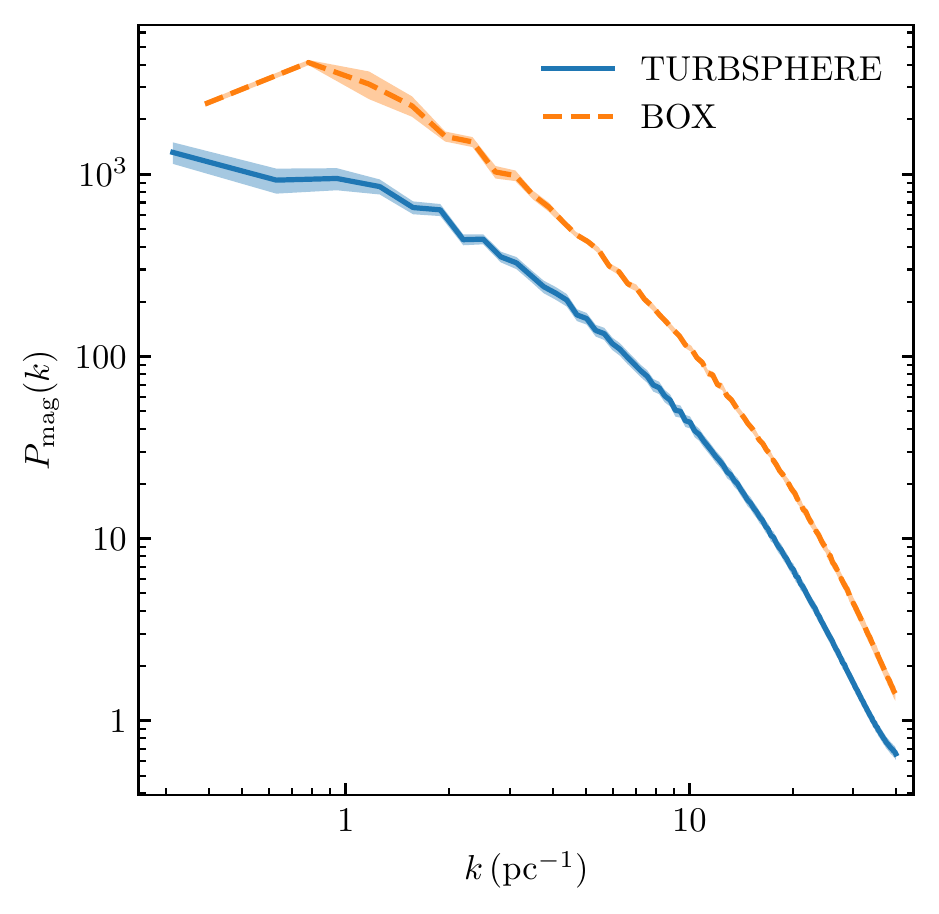}
    \vspace{-8mm}
    \caption{The magnetic field power spectrum of the the fiducial TURBSPHERE run, expressed as power spectrum of the magnetic field for given physical wavenumbers $k (\rm pc^{-1})$, compared to that of the equivalent BOX run. We plot the average statistics of the last five crossing times and their standard deviations (shaded regions).}
    \label{fig:powerspec_b}
\end{figure}

We also plot the magnetic field power spectrum $P_{\rm mag} \left(k\right)$ in Figure {\ref{fig:powerspec_b}}. The power spectra of the two setups are again qualitatively similar, with {\small TURBSPHERE} runs having a similar-yet-shallower slope for all values of $k$, which is opposite of what is seen in the velocity power spectra. In all, the basic density, velocity, and magnetic field statistics of the {\small TURBSPHERE} setup are fairly similar -- but not identical -- to those found in the {\small BOX} setup. Therefore, the turbulent properties of our {\small TURBSPHERE} setup demonstrates that this new method accurately models supersonic turbulence in GMCs for star formation simulations. 

\subsubsection{Sensitivity to parameters} 
We note the equilibrium values of various statistics ($\mathcal{M}_{\rm RMS}$, $R_{\rm 50}$, $R_{\rm RMS}$, $E_{\rm mag}/E_{\rm kin}$, and $\alpha_{\rm turb}$) depend on parameters such as the forcing strength scaling factor $\tilde{f}$, driving wavelength range scaling factor $\tilde{\lambda}$, analytic gravity well scaling factor $M_{\rm well}/M_{\rm 0}$, and initial mass-to-flux ratio $\mu_{\rm 0}$ in the {\small TURBSPHERE} stirring runs, listed in Table \ref{table:stirring_runs}. Unweighted least-squares logarthimic fits yield the following power-law relations that approximate all of our results to within $\sim 0.1 \rm dex$:
\begin{equation} \label{eq:r50}
    \frac{R_{\rm 50}}{R_{\rm 0}} \approx 0.82 \tilde{f}^{0.5} \tilde{\lambda}^{0.5} \left(\frac{M_{\rm well}}{M_{\rm 0}}\right)^{-0.5},
\end{equation}
\begin{equation} \label{eq:rrms}
    \frac{R_{\rm RMS}}{R_{\rm 0}} \approx 0.97 \tilde{f}^{0.6} \tilde{\lambda}^{0.8} \left(\frac{M_{\rm well}}{M_{\rm 0}}\right)^{-0.6},
\end{equation}
\begin{equation} \label{eq:eratio}
    \frac{E_{\rm mag}}{E_{\rm turb}} \approx 1.29 \tilde{f}^{0.4} \tilde{\lambda}^{-1} \mu_{\rm 0}^{-0.8} \left(\frac{M_{\rm well}}{M_{\rm 0}}\right)^{-0.3},
\end{equation}
\begin{equation} \label{eq:alpha}
    \alpha_{\rm turb}\approx  \tilde{f}^{1.5} \tilde{\lambda}^{1.6} \left(\frac{M_{\rm well}}{M_{\rm 0}}\right)^{-0.7},
\end{equation}

\begin{equation} \label{eq:mach}
    \mathcal{M} \approx 11 \tilde{f}^{0.4} \tilde{\lambda}^{0.4}.
\end{equation}

The above approximations neglect all powers shallower than $0.1$, and at this level of significance the only quantity that is sensitive to $\mu_{\rm 0}$ is the magnetic energy itself, while the Mach number $\mathcal{M}$ is insensitive to $\mu_{\rm 0}$ {\it and} the properties of the confining potential. Using these formulae, an initial condition with any desired bulk equilibrium properties may be generated (within their respective domains of validity).

\label{sec:sensitivity_to_params}

\subsection{Star formation runs}

\begin{figure*}
    \centering
    \begin{subfigure}[b]{0.2\textwidth}
        \includegraphics[width=\textwidth, height=12mm, keepaspectratio]{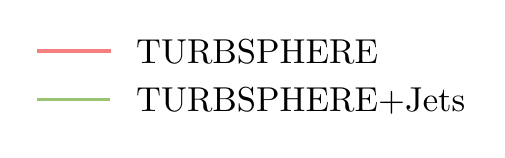}
        \vspace{-6mm}
    \end{subfigure}
    \begin{subfigure}[b]{0.26\textwidth}
        \includegraphics[width=\textwidth, height=25mm, keepaspectratio]{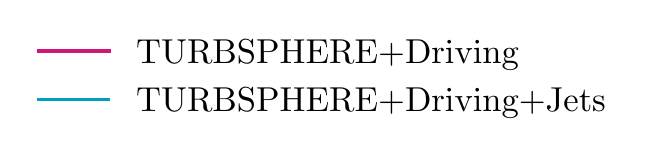}
        \vspace{-6mm}
     \end{subfigure}
    \begin{subfigure}[b]{0.2\textwidth}
        \includegraphics[width=\textwidth, height=12mm, keepaspectratio]{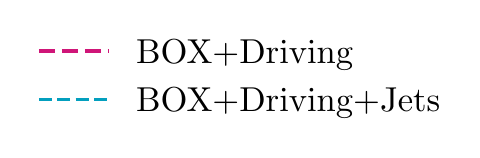}
        \vspace{-6mm}
    \end{subfigure}
    \begin{subfigure}[b]{0.18\textwidth}
        \includegraphics[width=\textwidth, height=12mm, keepaspectratio]{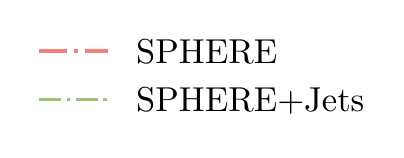}
        \vspace{-3mm}
    \end{subfigure}
    \begin{subfigure}[b]{\textwidth}
        \includegraphics[width=\textwidth]{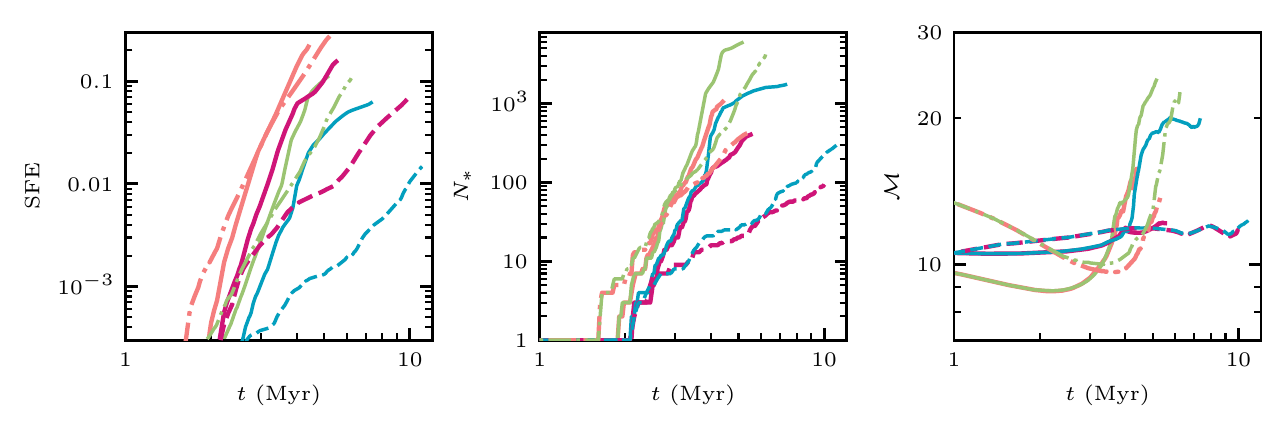}
        \vspace{-6mm}
    \end{subfigure}

    \caption{Statistics of the various {\small TURBSPHERE}, {\small BOX}, and {\small SPHERE} star formation runs with combinations of {\small Jets} and {\small Driving}. The initial condition setup of the {\small TURBSPHERE} runs are discussed in \S\ref{sec:star_formation_runs}, with initial condition parameters from the runs in Table \ref{table:runs_dir}. Panels show (1) the star formation efficiency and (2) the number of stars formed, (3) and the 3D Mach number $\mathcal{M}$ as a function of time (Myr).} 
    \label{fig:SF}
\end{figure*}

\begin{figure}
    \centering
    \includegraphics[width=\columnwidth]{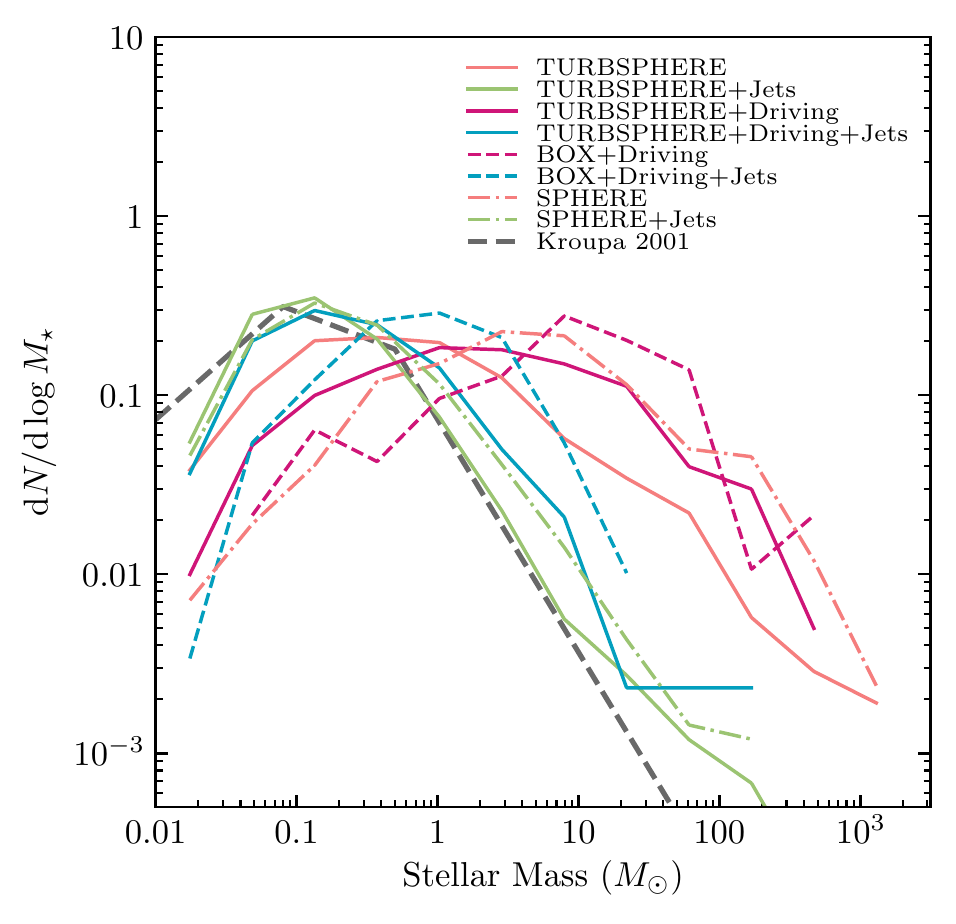}
    \vspace{-8mm}
    \caption{The stellar initial mass function (IMF) from the final snapshot of each star formation simulation with different physics and setups (see Table \ref{table:runs_dir}). We normalize each IMF and label the initial condition methods by linestyle and physics modules by colors. For comparison we plot the \citet{kroupa:imf} IMF.} 
    \label{fig:imf_plot}
\end{figure}

We now examine the results of the {\small BOX}, {\small SPHERE}, and {\small TURBSPHERE} simulations run with various combinations of physics (see Table \ref{table:runs_dir}). In Figure \ref{fig:SF} we compare the star formation histories of each run with their star forming {\small BOX} and {\small SPHERE} counterparts. For each simulation, we plot the star formation efficiency ($= \sum M_{\star}/M_{\rm 0}$), number of stars formed $N_{\star}$, and the 3D Mach number $\mathcal{M}$, as a function of time. 

We find general trends in star formation histories that differ between {\small BOX}, {\small SPHERE}, and {\small TURBSPHERE} simulations. {\small BOX} runs consistently show lower star formation efficiencies at fixed time, while {\small SPHERE} \& {\small TURBSPHERE} have generally more-similar SFE. Furthermore, the slope of the SFE time evolution in {\small BOX} runs is shallower (SFE $\propto (t-t_{0})^2$, e.g. \citealt{murray.chang:2015,lee:2015.gravoturbulence,murray:2017}), while {\small SPHERE} \& {\small TURBSPHERE} runs have steeper SFE slopes (similar to the $\propto (t-t_{0})^3$ law found in previous {\small SPHERE} runs, e.g. \citealt{guszejnov_isothermal_mhd,starforge_jets_imf}). A similar trend is also reflected in the number of stars $N_{\star}$, where {\small BOX} simulations exhibit a shallower SF history than the others. We further discuss the physical explanation of the faster star formation rate (SFR) in \S\ref{sec:initcondit}. 

Through the duration of the {\small BOX} simulations, we find a general stability in the Mach number. {\small TURBSPHERE+Driving} runs follow this general-stability trend, with divergence dependent on the presence of protostellar jets. In contrast, turbulence in the {\small SPHERE} runs decays quickly until feedback begins to inject significant kinetic energy into the cloud, which helps sustain turbulence \citep{Rosen2020a, Appel2021a}. {\small TURBSPHERE} runs without driving also follow this trend, with Mach numbers initially declining until feedback replenishes turbulence. The effects of driving and the Mach number are further discussed in \S\ref{sec:driving_effects}.
\label{sec:star_formation_runs}

We plot the stellar mass functions (specifically, the quantity $\mathrm{d}N/\mathrm{d} \log M_\star$) measured at the end of each simulation in Figure \ref{fig:imf_plot}. As found previously in {\small BOX} and {\small SPHERE} simulations \citep{guszejnov_isothermal_mhd}, {\small TURBSPHERE} simulations without any feedback result in a mass distribution that is much too shallow top-heavy, with most stellar mass contained in stars more massive than $20 M_\odot$. We do note some diversity in the shapes of the IMFs among the runs without feedback: the {\small TURBSPHERE} run peaks at a factor of $3-4$ lower than the {\small TURBSPHERE+Driving} run, and the {\small BOX+Driving} and {\small SPHERE} run mass functions peak at even greater masses. 

In agreement with prior work, we find protostellar jets reduce both the mass-scale of stars and the overall star formation efficiency  \citep{hansen:2012.jets, krumholz:2012.orion.rhd, cunningham:2018.protostellar.mhd, Rosen2020a, starforge_jets_imf, mathew:2021.imf.jets}. As shown in \citetalias{starforge_jets_imf}, protostellar jets interrupt the accretion of material onto already-existing stars, while also promoting fragmentation and the birth of low-mass stars. This impact is clearly shown in the IMF of the star formation runs in Figure \ref{fig:imf_plot}. In comparison to their counterparts, star-forming runs that include protostellar jet feedback promote significantly higher quantities of low-mass stars, resulting in better agreement with the observed IMF. 

The influence of protostellar jets upon the cloud kinematics is also heavily dependent on the star formation history, which varies in the different setups.  
This is most clear when comparing the SFE and $\mathcal{M}$ histories of the {\small TURBSPHERE+Driving+Jets} run to the {\small BOX+Driving+Jets} run, and their IMFs (Figures \ref{fig:SF} \& \ref{fig:imf_plot}, respectively).
While the IMFs of these two runs are qualitatively similar in shape, their final Mach numbers diverge. For the respective {\small BOX} run, the relative lack of stars results in far fewer protostellar jets and thus less energy injection into the system. While $\mathcal{M}$ is sustained throughout the simulation (due to continual energy injection via our driving scheme), the smaller number of stars and their protostellar jets results in a reduced ability to stir turbulence, as evidenced by a stagnant $\mathcal{M}$ in the {\small BOX} run. However, the more rapid star formation of the {\small TURBSPHERE} run allows for significant energy injection, allowing the Mach number to rise significantly. We can attribute this directly to the protostellar jets by comparing the {\small TURBSPHERE+Driving} run to the {\small TURBSPHERE+Driving+Jets} run, as the lack of protostellar jets results in an equilibrium Mach number being sustained, diverging from the run with protostellar jets enabled.

Lastly, although the IMF appears closer to the canonical \citet{kroupa:imf} shape when protostellar jets are enabled in {\small TURBSPHERE} runs, there is still a residual excess of high-mass ($\gtrsim 10 M_\odot$) stars seen in previous {\small SPHERE} simulations with protostellar jets \citepalias{starforge_jets_imf}. However, these runs do not include other feedback mechanisms such as radiation pressure, photoionization, and stellar winds, which are likely important for limiting the accretion of massive stars.

\label{sec:protostellar_jets} 

\section{Discussion}
\label{sec:discussion}

\subsection{Do initial conditions matter?}

To determine the influence of initial conditions while holding boundary conditions fixed, we compare the decaying {\small SPHERE} and {\small TURBSPHERE} runs in Figure \ref{fig:SF}, both with and without protostellar jet feedback. We note a delay in initial star formation in panel 1 for {\small TURBSPHERE} runs, where the {\small SPHERE} run begins forming stars approximately 1 Myr earlier. Despite this delay, the eventual star formation rate of {\small TURBSPHERE} is higher than that of {\small SPHERE}, resulting in {\small TURBSPHERE} overtaking {\small SPHERE} in SFE as time progresses. This is further supported in panel two, where the number of stars, $N_{\star}$, diverges, favoring {\small TURBSPHERE} as time progresses. We attribute these differences solely to the different initial conditions of the simulation.

As the {\small TURBSPHERE} simulation begins with self-consistent MHD turbulence, the magnetic field has already been amplified by the magnetic dynamo, and thus the gas initially has more magnetic pressure support, which has been shown to suppress star formation by reducing turbulent fragmentation \citep[e.g.,][]{Rosen2020a, Appel2021a}. However, due to the realistic MHD turbulence of {\small TURBSPHERE}, pre-assembled turbulent structures, such as dense cores and filaments, are permitted to form \emph{before} stars can form. Such cores provide pockets of dense gas where the Jeans mass is lower, encouraging fragmentation. As such structures form earlier in {\small TURBSPHERE}, star formation is encouraged at higher rates in comparison to {\small SPHERE} runs, once it begins. To a lesser degree, these phenomena (of delayed star formation \& increased SFR in {\small TURBSPHERE} runs) are found when comparing {\small TURBSPHERE+Jets} and {\small SPHERE+Jets} runs.

To assess the impact of initial conditions on the qualitative structure of the IMF, we compare {\small TURBSPHERE} and {\small SPHERE} runs without jets or driving in Figure \ref{fig:imf_plot}. We find that the IMF turnover (where the power-law transitions to a flat segment) for the {\small TURBSPHERE} run occurs at slightly lower stellar masses than {\small SPHERE} runs. Compared to {\small SPHERE}, the {\small TURBSPHERE} setup produces higher quantities of low-mass stars for both runs with and without protostellar jets, however the difference in the shape of the IMF appears is much less pronrounced with feedback.

Note that the IMFs of these runs were taken from the final snapshot of our simulations. As a result, the final SFE of our clouds differs between the plotted IMFs in Figure \ref{fig:imf_plot} (final SFEs are listed in Table \ref{table:runs_dir}). We do not take the IMF at fixed SFE as {\small BOX} runs have a significantly suppressed SFE, making comparison to {\small SPHERE} and {\small TURBPSHERE} at fixed SFE difficult. Thus, we focus on the IMF turnover as our IMFs are likely more top-heavy than physically realistic. A more conclusive analysis exploring the origins of these differences requires a larger statistical sample of both {\small SPHERE} and {\small TURBSPHERE} setups with full-physics runs.

\label{sec:initcondit}

\subsection{Disentangling the roles of driving and boundary conditions}
Simulations have historically varied {\it both} turbulence driving and boundary conditions together between the {\small SPHERE} and {\small BOX} setups, as a result it has been difficult to attribute various differences in star formation results to one factor or the other. Our numerical experiments allow us to disentangle these two effects.

\subsubsection{Boundary conditions}
To isolate the role of periodic and non-periodic boundary conditions on the results, we compare the {\small BOX} and {\small TURBSPHERE+Driving} runs in Figure \ref{fig:SF}. Compared to the {\small TURBSPHERE} runs, we find significant reduction of the SFE and quantity of stars in {\small BOX} runs, with the SFE exhibiting a significantly shallower slope with respect to simulation time. This same SFE suppression  has been noted previously between {\small SPHERE} and {\small BOX} setups in studies that have directly compared the two under controlled conditions \citep{krumholz:2012.orion.rhd, guszejnov_isothermal_mhd} and has been broadly attributed simply to the moderation of star formation by  turbulence \citep[e.g.][]{km2005,federrath_klessen_2012}. However, our results show that the boundary conditions also contribute significantly.

To investigate the role of boundary conditions on the IMF, we compare the results of {\small TURBSPHERE} and {\small BOX} runs in Figure \ref{fig:imf_plot}. We find that {\small BOX} produces relatively lower amounts of low-mass stars for runs with and without jets. We note that the IMF turnover skews toward higher masses in the {\small BOX} runs (as was also found in \citetalias{starforge_jets_imf}). We attribute the difference in the shift of the IMF turnover soley to the different boundary conditions of {\small TURBSPHERE} and {\small BOX}.

Differences in both the SF history and stellar mass statistics are likely explained by the importance of global infall in the {\small TURBSPHERE} (and {\small SPHERE}) setups. As discussed in \S\ref{sec:intro}, for a given set of turbulent initial conditions the {\small BOX} setup generally has about an order of magnitude less gravitational energy due to the different boundary conditions of the Poisson equation. Consequently, {\small BOX} runs effectively have a much higher virial parameter \citep{federrath_klessen_2012}, inhibiting large-scale gravitational collapse. Conversely, both {\small SPHERE} and {\small TURBSPHERE} clouds are subject to their own global weight, and as a result tend to become centrally concentrated over time, despite driving. This may funnel gas to higher densities, resulting in more low-mass stars (due to a reduced Jeans mass) and encouraging star formation overall. 
\label{sec:boundaryconditions}

\subsubsection{Effects of driving} 
The inclusion of turbulent driving in some of our simulations models large-scale energy injection produced by the larger galactic environment. To isolate the influence of driving, we primarily compare {\small TURBSPHERE} runs with and without {\small Driving} enabled as we have a controlled experiment where non-driving factors are held fixed. We find that turbulent driving reduces the SFE and the number of stars at the same time interval. Lower SFEs and star counts are observed both with and without protostellar jets. We conclude that turbulent driving moderates star formation in its own right, as shown in previous works that aimed to disentangle various factors in a {\small BOX}-like setup \citep{federrath_2015_inefficient_sf}.

Interestingly, driving in concert with protostellar jets appears to {\it quench} star formation in the cloud at an apparent final SFE approaching $\sim 7\%$. The {\small SPHERE} and undriven {\small TURBSPHERE} run with jets also show some evidence of the SFE tapering off (similar behaviour was also found in \citetalias{starforge_jets_imf}) but at significantly higher ($\gtrsim 15\%$) SFE. So in concert with feedback, driving can also reduce the {\it final} SFE of a GMC, presumably by providing pressure support (or in other words: increasing the virial parameter), which makes the cloud easier to disrupt. We note that the {\small TURBSPHERE} and {\small SPHERE} driving runs exhibit a top-heaviness in comparison to their non-driving counterparts in their IMFs. Since the growth of median and maximum stellar mass tends to correlate strongly with the SFE in simulations \citep{bonnell:2003.hierarchical, peters:2010.sf.rhd.sims,  guszejnov_isothermal_mhd}, earlier termination of star formation can be expected to moderate the top-heaviness of the IMF. 
\label{sec:driving_effects}

\section{Conclusion}
\label{sec:conclusion}
We have demonstrated a new setup for GMC-scale star formation simulations -- called {\small TURBSPHERE} -- that simultaneously models the self-consistent MHD turbulence, large-scale energy injection that realistically sustains large-scale turbulence, and spatial concentration of real GMCs. This new setup combines the advantages of previous {\small SPHERE} and {\small BOX} setups that are commonly used for simulating star formation that occurs within GMCs. Through the introduction of an analytic gravity well and continuous driving, we generated an isolated turbulent cloud in dynamical equilibrium (Figures \ref{fig:multi_panel} and \ref{fig:evolution}). From our suite of turbulence driving runs, we find: 
\begin{itemize}
    \item The {\small TURBSPHERE} initial condition has properties resembling those found in the well-studied periodic {\small BOX} setup, as evidenced by the similarity in the density PDF and the velocity and magnetic field power spectra. We note some slight differences in the spectral indices and large-scale density structure, which we discuss in \S\ref{sec:stirring_runs_sec}. (Figures \ref{fig:rhopdf}, \ref{fig:powerspec_v}, and \ref{fig:powerspec_b}).
    
    \item We map out the relations between the normalized forcing strength scale factor $\tilde{f}$, driving wavelength range scale factor $\tilde{\lambda}$, analytic gravity well scaling factor $M_{\rm well}/M_{\rm 0}$, and initial mass-to-flux ratio $\mu_{\rm 0}$ for generating various equilibrium statistics ($\mathcal{M}_{\rm RMS}$, $R_{\rm 50}$, $R_{\rm RMS}$, $E_{\rm mag}/E_{\rm kin}$, and $\alpha_{\rm turb}$) for particular initial conditions generated by the {\small TURBSPHERE} setup. These power-law relations are listed in Equations \ref{eq:r50}, \ref{eq:rrms}, \ref{eq:eratio}, \ref{eq:alpha}, and \ref{eq:mach}, and discussed in \S\ref{sec:sensitivity_to_params}.
\end{itemize}
We then performed a suite star formation simulations, with and without protostellar jets and driving, with various initial conditions (listed in Table \ref{table:runs_dir}). These results make it possible to isolate and study the effects of the turbulent cascade, simulation geometry, external driving, and gravity/MHD boundary conditions upon the star formation history and IMFs predicted by simulations. We plot these quantities in Figures \ref{fig:SF} and \ref{fig:imf_plot} and found: 
\begin{itemize}
   
    \item The periodic boundary conditions of the {\small BOX} setup suppress gravitational collapse of the cloud, resulting in a significantly shallower SF history than the {\small TURBSPHERE} run with equivalent turbulent conditions and driving. The SF history of the {\small TURBSPHERE} simulations obeys a similar steep ($\mathrm{SFE} \sim  t^3$) law as previously seen in {\small SPHERE} simulations, regardless of driving or jet feedback. (\S\ref{sec:boundaryconditions})

    \item Whether in the {\small TURBSPHERE} or {\small BOX} setup, and with or without feedback, the continued driving of turbulence moderates star formation significantly, with or without protostellar jets. Driven runs have a SFE that is a factor of $\sim 2-3$ lower at fixed times compared to their undriven counterparts. Driving in concert with jets can disrupt the GMC sufficiently to nearly halt star formation with an SFE $ \lesssim 10\%$. (\S\ref{sec:driving_effects}) 

    \item Without feedback , and with or without driving, the stellar mass function in {\small TURBSPHERE} runs remains much shallower and more top-heavy than observations, in agreement with previous conclusions from {\small BOX} and {\small SPHERE} runs \citep{guszejnov_isothermal_mhd}. However, the detailed shape of the mass function does differ in each different setup, with {\small TURBSPHERE} having the lowest IMF peak mass of the three.
    
    \item With protostellar jet feedback, {\small TURBSPHERE} runs produce far more low-mass stars and obtain a mass function more more similar to e.g. the \citet{kroupa:imf} form, again in agreement with previous results from the other setups \citepalias[e.g.][]{starforge_jets_imf}. However a shallower slope at very high masses ($>20 M_\odot$) remains, indicating that additional physics are required to regulate the growth of the most massive stars. 

\end{itemize}
We find that {\small TURBSPHERE} offers a more realistic approach to generating the initial conditions of isolated GMCs and modeling their subsequent gravitational collapse that drives star formation within them, however we note some caveats remain since GMCs are not isolated objects. It may be important to model the formation of GMCs within galactic contexts, notably through large-scale colliding flows, cloud-cloud collisions, and thermal and gravitational instability \citep{Tasker:2009.gmc.collisions.disk, Wu_2017,zamora:2019.grav.feedback, Chevance:2020.gmc.lifecycle}, which the {\small TURBSPHERE} setup does not model. Furthermore, while the difference between the analytic potential and the eventual gas self-gravity is small by construction, it is still unphysical to instantly "switch on" self-gravity.

The {\small TURBSPHERE} setup permits simulations with turbulence driving in concert with all feedback mechanisms, including radiation, stellar winds, and supernovae, in addition to the protostellar jets considered here. Such simulations will allow a more in-depth study of the importance of large-scale energy injection on the star formation rate and the IMF in concert with feedback, bringing us closer to understanding star formation in GMCs in a galactic context. The results of these simulations and their analysis will be presented in future works. 

\section*{Acknowledgements}
 Support for MYG and HBL was provided by a CIERA Postdoctoral Fellowship. Support for MYG was also provided by NASA through the NASA Hubble Fellowship grant \#HST-HF2-51479 awarded  by  the  Space  Telescope  Science  Institute,  which  is  operated  by  the   Association  of  Universities  for  Research  in  Astronomy,  Inc.,  for  NASA,  under  contract NAS5-26555. This work used computational resources provided by Frontera allocation AST20019 and AST21002, and additional resources provided by the University of Texas at Austin and the Texas Advanced Computing Center (TACC; http://www.tacc.utexas.edu). This research is part of the Frontera computing project at the Texas Advanced Computing Center. Frontera is made possible by National Science Foundation award OAC-1818253. DG was supported by the Harlan J. Smith McDonald Observatory Postdoctoral Fellowship and the Cottrell Fellowships Award (\#27982) from the Research Corporation for Science Advancement. SSRO was supported by NSF CAREER grant AST-1748571. CAFG was supported by NSF through grants AST-1715216, AST-2108230,  and CAREER award AST-1652522; by NASA through grant 17-ATP17-0067; by STScI through grant HST-AR-16124.001-A; and by the Research Corporation for Science Advancement through a Cottrell Scholar Award. ALR  acknowledges support from Harvard University through the ITC Post-doctoral Fellowship.

\section*{Data availability}
The data supporting the plots within this article and the initial conditions used for numerical tests are available upon request to the corresponding authors. A public version of the {\small GIZMO} code is available at \url{http://www.tapir.caltech.edu/~phopkins/Site/GIZMO.html}.



\bibliographystyle{mnras}
\bibliography{bibliography} 








\bsp	
\label{lastpage}
\end{document}